\documentclass[twocolumn]{aastex6}

\newcommand{\about}{$\sim$~}

\newcommand{\degrees}{$^{\circ}$}

\newcommand{\asec}{$\arcsec$}
\newcommand{\fasec}{$\farcs$}
\newcommand{\lprime}{$L^{\prime}$}

\newcommand{\microns}{$\mu$m}
\newcommand{\msun}{M$_{\odot}$}
\newcommand{\ms}{m s$^{-1}$}

\newcommand{\reducedchi}{$\chi_{\nu}^{2}$}

\begin{document}


\title{MagAO Imaging of Long-period Objects (MILO). II. A Puzzling White Dwarf around the Sun-like Star HD 11112}

%% Use \author, \affil, plus the \and command to format author and affiliation 
%% information.  If done correctly the peer review system will be able to
%% automatically put the author and affiliation information from the manuscript
%% and save the corresponding author the trouble of entering it by hand.
%%
%% The \affil should be used to document primary affiliations and the
%% \altaffil should be used for secondary affiliations, titles, or email.

%% Authors with the same affiliation can be grouped in a single

\author{Timothy J. Rodigas\altaffilmark{1,13}, P. Bergeron\altaffilmark{2}, Am{\'e}lie Simon\altaffilmark{2}, Pamela Arriagada\altaffilmark{1}, Jackie Faherty\altaffilmark{1}, Guillem Anglada-Escud{\'e}\altaffilmark{3}, Eric E. Mamajek\altaffilmark{4}, Alycia Weinberger\altaffilmark{1}, R. Paul Butler\altaffilmark{1}, Jared R. Males\altaffilmark{5}, Katie Morzinski\altaffilmark{5}, Laird M. Close\altaffilmark{5}, Philip M. Hinz\altaffilmark{5}, Jeremy Bailey\altaffilmark{6,7}, Brad Carter\altaffilmark{8}, James S. Jenkins\altaffilmark{9}, Hugh Jones\altaffilmark{10}, Simon O'Toole\altaffilmark{11}, C.G. Tinney\altaffilmark{6,7}, Rob Wittenmyer\altaffilmark{6,7}, John Debes\altaffilmark{12}}

\altaffiltext{1}{Department of Terrestrial Magnetism, Carnegie Institution of Washington, 5241 Broad Branch Road, NW, Washington, DC 20015, USA; email: trodigas@carnegiescience.edu}
\altaffiltext{2}{D{\'e}partement de Physique, Universit{\'e} de Montr{\'e}al, C.P.~6128, Succ.~Centre-Ville, Montr{\'e}al, Qu{\'e}bec H3C 3J7, Canada}
\altaffiltext{3}{School of Physics and Astronomy, Queen Mary, University of London, 327 Mile End Rd. London, UK}
\altaffiltext{4}{Department of Physics and Astronomy, University of Rochester, Rochester, NY 14627-0171, USA}
\altaffiltext{5}{Steward Observatory, The University of Arizona, 933 N. Cherry Ave., Tucson, AZ 85721, USA}
\altaffiltext{6}{Exoplanetary Science at UNSW, School of Physics, UNSW Australia, Sydney, NSW 2052, Australia}
\altaffiltext{7}{Australian Centre for Astrobiology, UNSW Australia, Sydney, NSW 2052, Australia}
\altaffiltext{8}{Computational Engineering and Science Research Centre, University of Southern Queensland, Springfield, QLD 4300, Australia}
\altaffiltext{9}{Departamento de Astronomia, Universidad de Chile, Casilla 36-D, Las Condes, Santiago, Chile}
\altaffiltext{10}{Centre for Astrophysics Research, University of Hertfordshire, College Lane, Hatfield, Herts AL10 9AB, UK}
\altaffiltext{11}{Australian Astronomical Observatory, PO Box 915, North Ryde, NSW 1670, Australia}
\altaffiltext{12}{Space Telescope Science Institute, Baltimore, MD 21218, USA}
\altaffiltext{13}{Hubble Fellow}

\begin{abstract}
HD 11112 is an old, Sun-like star that has a long-term radial velocity (RV) trend indicative of a massive companion on a wide orbit. Here we present direct images of the source responsible for the trend using the Magellan Adaptive Optics system. We detect the object (HD 11112B) at a separation of 2\fasec 2 (100 AU) at multiple wavelengths spanning 0.6-4 \microns ~and show that it is most likely a gravitationally-bound cool white dwarf. Modeling its spectral energy distribution (SED) suggests that its mass is 0.9-1.1 \msun, which corresponds to very high-eccentricity, near edge-on orbits from Markov chain Monte Carlo analysis of the RV and imaging data together. The total age of the white dwarf is $>2\sigma$ discrepant with that of the primary star under most assumptions. The problem can be resolved if the white dwarf progenitor was initially a double white dwarf binary that then merged into the observed high-mass white dwarf. HD 11112B is a unique and intriguing benchmark object that can be used to calibrate atmospheric and evolutionary models of cool white dwarfs and should thus continue to be monitored by RV and direct imaging over the coming years.
\end{abstract}

\keywords{instrumentation: adaptive optics --- techniques: high angular resolution --- techniques: radial velocity --- stars: individual (HD 11112) --- binaries --- white dwarfs} 

\section{Introduction} \label{sec:intro}
Direct imaging and Doppler spectroscopy are complementary techniques for characterizing planetary systems. The former can detect young, massive companions on wide orbits, while the latter is most sensitive to massive companions orbiting close to their typically old, chromospherically-quiet host stars. The combination of the two techniques has now been exploited in several large programs: the NACO-SDI survey \citep{jenkinsrvimaging}, the TRENDS survey \citep{crepptrends1,crepptrends2,crepptrends3,crepptrends4,crepptrends5,crepptrends6}, the Friends of Hot Jupiters survey \citep{friendsofhotjupiters1,friendsofhotjupiters2}, and the Subaru/HiCIAO survey \citep{subarutrends}. In addition to these, for the past few years we have been executing our own survey, MagAO Imaging of Long-period Objects (MILO), which uses the superb visible and near-infrared (NIR) imaging capabilities of the Magellan adaptive optics (MagAO; \citealt{lairdtrapezium}) system in combination with precision radial velocities (RVs) to discover and characterize wide companions. In our first paper, we described the discovery and characterization of a benchmark mid-M dwarf (HD 7449B) that is likely to be inducing Kozai oscillations on a very nearby gas giant planet (HD 7449Ab) \citep{milo1}. 

In this paper, we present the discovery and characterization of a faint white dwarf orbiting the Sun-like star HD 11112. This star, located 45.3$^{+1.2}_{-1.1}$ pc away \citep{updatedhip}, has a spectral type of \about G2 (ranging from G0-G4; \citealt{sptypesbook,evans,bidelman}), is metal-rich ($[Fe/H] = 0.20\pm 0.06$, \citealt{bensby}), is thought to be old (\about 4-8 Gyr; \citealt{valenti,bensby,ghezziages,holmberg,ramirezlithium,feltzing}) based on its chromospheric activity and kinematics, and is likely evolving off the main sequence. The star has been monitored for the past 17 years by the Anglo-Australian Telescope (AAT) UCLES spectrometer, revealing a long-term linear trend indicative of a massive companion on a wide orbit. In Section 2, we describe our high-contrast imaging and Doppler spectroscopy observations and data reduction. In Section 3, we present our astrometry and photometry of the directly imaged companion, model its spectral energy distribution (SED) using cool white dwarf model atmospheres, and constrain its mass via analysis of the RVs. In Section 4, we summarize and discuss the nature of this puzzling companion based on all the information at hand on the system. 

\section{Observations and Data Reduction} 
\subsection{MagAO Imaging} \label{sec:obsmagao}
We observed HD 11112 using the Magellan Clay telescope at Las Campanas Observatory in Chile on the nights of UT November 8-10, 2014 (epoch 1), and \about 1 year later on the night of UT November 30, 2015 (epoch 2). Observations were made using both the visible camera VisAO \citep{visao} and the infrared camera Clio-2 \citep{suresh}. VisAO has a plate scale of 0\fasec 0079 and a field of view (FOV) of \about 8\asec ~\citep{jaredbetapic}. For Clio-2, we used the narrow camera, which has a plate scale of 0\fasec 01585 and a FOV of \about 9$\times$15\asec ~\citep{ktbetapic}. The instrument rotator was turned off for both cameras to enable angular differential imaging (ADI; \citealt{adi}). We observed in the $r'$ (0.63 \microns), $i'$ (0.77 \microns), $z'$ (0.91 \microns), $Ys$ (0.99 \microns), $J$ (1.1 \microns), H (1.65 \microns), $Ks$ (2.15 \microns), and \lprime ~(3.76 \microns) filters, detecting a faint point-source \about 2.2\asec ~away from the host star at a position angle of \about 226\degrees ~in all images. Unsaturated photometric images were acquired after each imaging sequence; for VisAO at $r'$ and $i'$, due to the brightness of the star, a neutral density (ND) filter was required to obtain unsaturated calibration images. For epoch 1, the observing conditions were fair, with variable seeing, strong wind, and intermittent cirrus clouds throughout both nights. For epoch 2, the conditions were much better, with clear skies and steady seeing under 1\asec ~for most of the night. See Table \ref{tab:obs} for a summary of all MagAO observations.
\begin{deluxetable*}{c|ccccc}
\tablecaption{Summary of MagAO Observations \label{tab:obs}}
\tablehead{
\colhead{UT Date} & \colhead{Camera} & \colhead{Filter} & \colhead{Total Exposure (min)} & Sky Rotation (degrees)\tablenotemark{a} & \colhead{Calibration\tablenotemark{b}} 
}
\startdata
2014 8 Nov. & VisAO & $z'$ & 11.75 & -- & Short \\
2014 9 Nov. & VisAO & $i'$ & 11.86 & -- & ND \\
2014 9 Nov. & Clio-2 & $J$ & 21 & 60.88 & Short \\
2014 9 Nov. & Clio-2 & H & 21 & 60.98 & Short \\
2014 9 Nov. & Clio-2 & $Ks$ & 19.13 & 52.34 & Short \\
2014 10 Nov.  & VisAO & $r'$ & 18.26 & -- & ND \\
2014 10 Nov. & Clio-2 & \lprime & 15.83 & 39.56 & Short \\
2015 30 Nov. & VisAO & $r'$ & 9.62 &  -- &  ND \\
2015 30 Nov. & VisAO & $i'$ & 8.87 & -- & ND \\
2015 30 Nov. & VisAO & $z'$ & 9.28 & -- & Short \\
2015 30 Nov. & VisAO & $Ys$ & 9.74 & -- & Short \\
2015 30 Nov. & Clio-2 & $J$ & 13.33 & 30.83 & Short \\
2015 30 Nov. & Clio-2 & H & 13.33 & 27.05 & Short \\
2015 30 Nov. & Clio-2 & $Ks$ & 13.33 & 19.85 & Short \\
\enddata
\tablenotetext{a}{Only relevant for ADI, which was not used in the reductions of the VisAO data.}
\tablenotetext{b}{Describes whether photometric images were acquired using an ND or unsaturated minimum exposures.}
%\tablecomments{blah}
\end{deluxetable*}

\begin{figure*}[t!]
%\figurenum{1}
\includegraphics[width=0.99\textwidth]{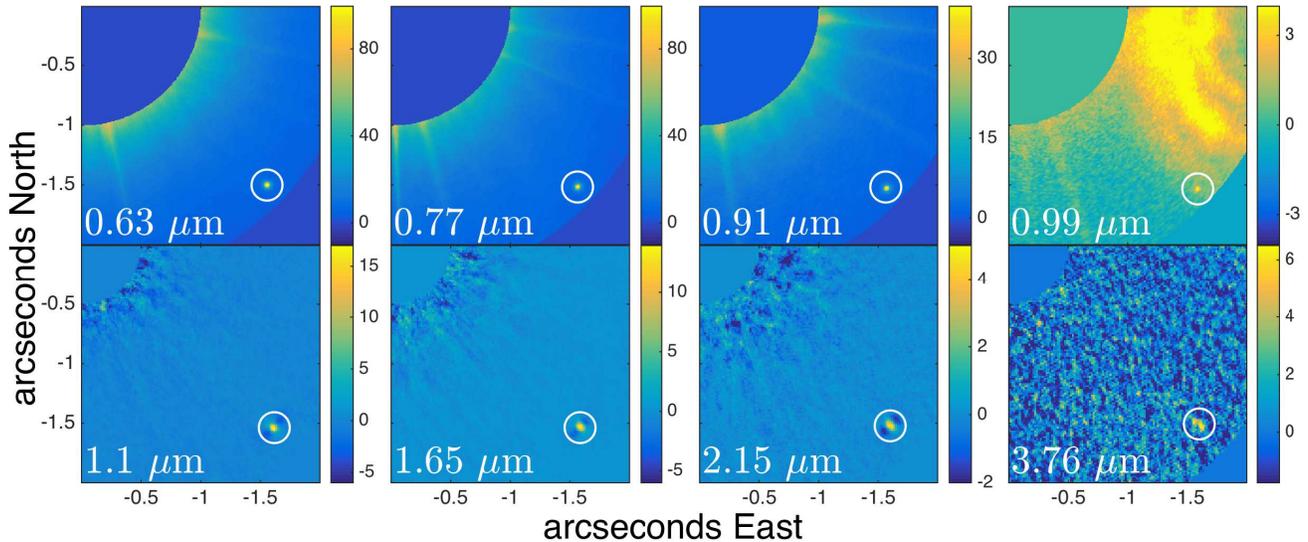}
\caption{MagAO images of the faint companion HD 11112B (from either epoch 1 or 2 depending on which had the highest SNR). North is up, east is to the left, and the primary star (HD 11112A) is at the top left corner of every image. Units are detector counts/s. The companion is circled and is located \about 2.2\asec ~away at a position angle of \about 226\degrees. Top row (from left to right): VisAO images at $r', i', z',$ and $Ys$, respectively. The $Ys$ image features a prominent reflection ghost above the companion and is therefore not used in the SED fitting. Bottom row (from left to right): Clio-2 images at $J, H, Ks,$ and \lprime. The \lprime ~detection is the only marginal detection, with SNR = 5.85.}
\label{fig:allimages}
\end{figure*}

The VisAO images were dark-subtracted, registered, divided by the integration times to give units of counts/s, cropped, and had their 2D radial profiles removed (since ADI was not used to reveal the companion). The images were then rotated to north-up, east-left and median-combined into final images (shown in Fig. \ref{fig:allimages}). The Clio-2 images were sky-subtracted, divided by the coadds and integration times to give units of counts/s, registered, cropped, and corrected for non-linearity \citep{ktbetapic}. Due to the sky brightness at 1-4 \microns, we used ADI and Principal Component Analysis (PCA, \citealt{pca}) to reveal HD 11112B. The number of PCA modes used at a given wavelength was determined by maximizing the SNR of the faint point-source. For epoch 1, the number of modes at $J, H, Ks,$ and \lprime ~was 6, 18, 10, and 4, respectively; for epoch 2, the number of modes at $J, H,$ and $Ks$ was 4, 7, and 4, respectively. After PCA PSF subtraction, the images were rotated to north-up, east-left and then median-combined into final images (shown in Fig. \ref{fig:allimages}). The unsaturated calibration images of the star were reduced in analogous ways for both VisAO and Clio-2.

\subsection{Doppler Spectroscopy} \label{sec:doppler}
\begin{figure}[h!]
\includegraphics[width=0.5\textwidth]{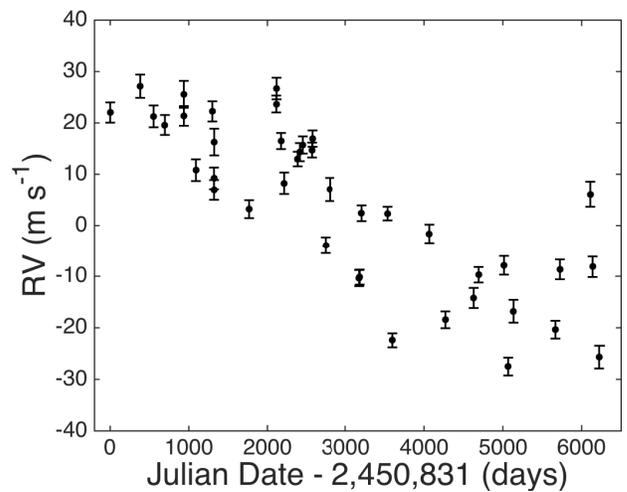}
\caption{RVs for HD 11112, obtained by the AAT/UCLES instrument over the last \about 17 years. There is no statistically significant curvature in the trend and no other apparent planetary signals.}
\label{fig:RVs}
\end{figure}

Observations of HD 11112 were obtained using the UCLES echelle spectrograph \citep{ucles} at the AAT by the Anglo-Australian Planet Search (AAPS) team. The observations for this star began in January 1998. The AAPS uses a 1\asec ~slit to obtain spectra with a resolution of $\sim$ 45,000 in the iodine region (5000-6300 \AA). A temperature-controlled iodine absorption cell \citep{iodine,pauliodine} is mounted in front of the instrument's entrance slit, imprinting the reference iodine spectrum directly on the incident starlight and providing a wavelength scale and measurement of the effective PSF for every observation \citep{pauliodine}. The median internal uncertainty achieved for HD 11112 using the iodine-cell technique with UCLES was 1.92 m $s^{-1}$. The RVs are reported in Table \ref{tab:rv} and shown in Fig. \ref{fig:RVs}.

\begin{deluxetable}{ccc}
\tablecaption{RVs for HD 11112 \label{tab:rv}}
\tablehead{
\colhead{Julian Date} & \colhead{RV (m s$^{-1}$)} & \colhead{$\sigma_{RV}$ (m s$^{-1}$)} 
}
\startdata
2450831.02725  &   22.02 &  1.97 \\
2451212.94298  &   27.15   &2.26 \\
2451383.30119   &  21.27   &2.13\\
2451526.99035    & 19.60   &1.94\\
2451767.29406     &25.57   &2.62\\
2451768.29660     &21.36   &1.91\\
2451920.98840     &10.79   &2.12\\
2452130.25817     &22.24   &1.96\\
2452151.24835      &9.19   &2.12\\
2452152.11663      &6.96   &1.92\\
2452154.24584     &16.28   &2.60\\
2452598.11249      &3.21   &1.75\\
2452943.12974     &23.68   &1.65\\
2452947.07961     &26.69   &2.10\\
2453003.96813     &16.47   &1.57\\
2453043.96771      &8.25   &2.08\\
2453216.32277     &12.97   &1.42\\
2453243.30378     &14.31   &1.76\\
2453281.16237     &15.70   &1.67\\
2453398.94062     &14.72   &1.45\\
2453404.97959     &16.94   &1.59\\
2453573.29648     &-3.90   &1.58\\
2453629.16886      &7.04   &2.25\\
2454009.18172    &-10.36  & 1.57\\
2454016.23685    &-10.17   &1.46\\
2454040.09766     & 2.41   &1.53\\
2454369.18469     & 2.34   &1.33\\
2454430.00539    &-22.44  & 1.36\\
2454898.90773     &-1.62   &1.81\\
2455102.18348    &-18.46   &1.63\\
2455461.15936    &-14.22   &1.96\\
2455524.04923     &-9.71   &1.51\\
2455846.11698     &-7.87   &1.82\\
2455899.03674    &-27.52   &1.73\\
2455966.91608    &-16.79   &2.22\\
2456498.30094    &-20.38   &1.72\\
2456555.18295     &-8.67   &1.93\\
2456940.15233      &6.11   &2.44\\
2456968.98567     &-8.16   &2.00\\
2457051.94494    &-25.69   &2.20\\
\enddata
\tablecomments{All reported RVs were obtained with AAT/UCLES.}
\end{deluxetable}

\section{Results}
\subsection{HD 11112 Stellar Properties}
\label{sec:stellar}
\begin{figure}[h!]
\centering
\includegraphics[width=0.48\textwidth]{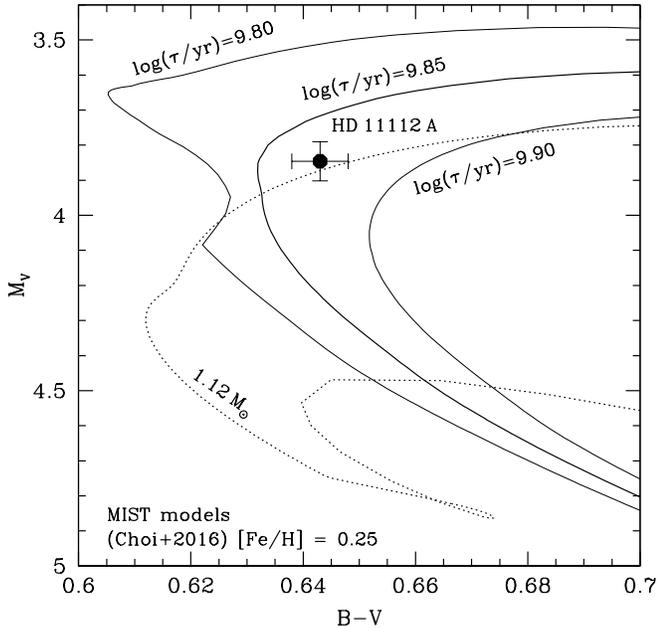}
\caption{Color-magnitude diagram with MIST tracks from \cite{mist}. Assuming $[Fe/H] = 0.25$ for HD 11112A, the star is consistent with having a mass of 1.12 \msun ~and age \about 7 Gyr.}
\label{fig:age}
\end{figure}

The age of HD 11112A has been estimated by several studies and ranges from \about 4-8 Gyr (\citealt{valenti,bensby,ghezziages,holmberg,ramirezlithium,feltzing}). The star's space velocity ($UVW$ from \citealt{holmberg}) points to an old age, as it is likely in the Hercules stream \citep{rayma}, and the star is chromospherically older than the Sun ($\log{R'HK}$ \about -5.0; \citealt{jenkinsactivity,paceactivity}). HD 11112A is likely slightly evolved, since its $\log g$ (\about 4.2; \citealt{ghezzilithium,bensby,valenti,ramirezlithium}) is smaller than what is typical for main sequence dwarfs ($\log g$ \about 4.3-4.5). Using the new MIST tracks with a metallicity of $[Fe/H] = 0.25$ (close to the metallicity of HD 11112A, $[Fe/H] = 0.20$, \citealt{bensby}), we find that HD 11112A is consistent with being a \about 1.12 \msun ~star with age \about 7 Gyr (see Fig. \ref{fig:age}). Thus multiple lines of evidence point to HD 11112A being old (likely older than the Sun) and to it evolving off the main sequence right now. We adopt as the age of HD 11112A the average of the previous measurements, excluding the three isochronal ages that are younger than the Sun \citep{valenti,bensby,ghezziages}; this leads to an age of 7.2$^{+0.78}_{-1.2}$ Gyr, consistent with our own new estimate using the MIST tracks (\about 7 Gry). 

\begin{figure}[h!]
\centering
\includegraphics[width=0.48\textwidth]{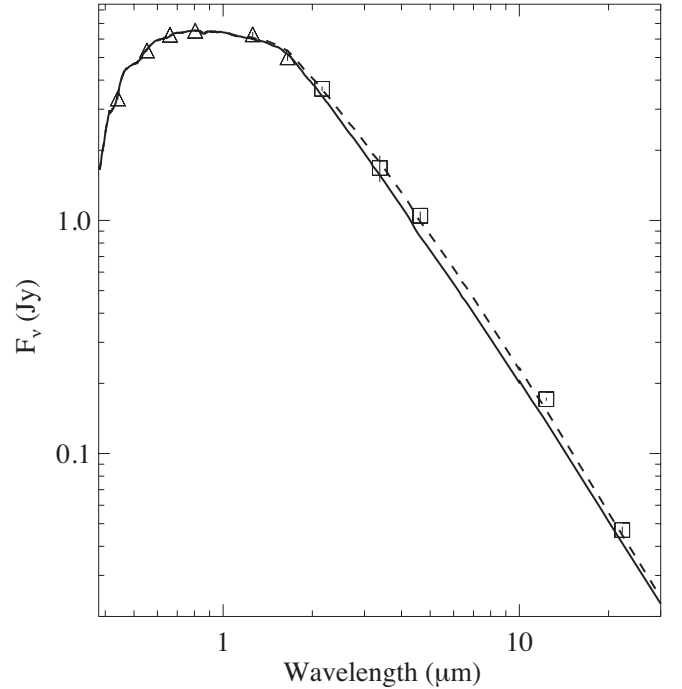}
\caption{SED of HD 11112A. The squares correspond to data that are in excess of the nominal stellar fit (solid curve). The excess (dashed curve) corresponds to a dust temperature of 1465 $K$ and $L_{ir}/L_{*}$ = 0.01 if the dust is around the primary. }
\label{fig:excess}
\end{figure}

The SED of HD 11112A, computed by comparing the star's apparent magnitudes at $B, V$ (converted from $BT$ and $VT$ from the \textit{Tycho} catalog, \citealt{tycho}), $R, I$ (from \citealt{cousinsphot}), $J, H, Ks$ (from \textit{2MASS}, \citealt{twomass}), and $W1, W2, W3$, and $W4$ (from \textit{ALLWISE}) is shown in Fig. \ref{fig:excess}. We fit the photometry to a grid of Kurucz/Castelli stellar models\footnote{\url{http://kurucz.harvard.edu/grids/gridP00ODFNEW/}} using different $\log{g}$ values (which made little difference). We also tested different $\log$ metallicity values relative to the Sun and found these also made little difference. The final best-fit model yielded an effective temperature of 6000 $K$ on the 250 $K$ temperature grid, which is consistent with previous measurements (\about 5900 from e.g., \citealt{bensby,ghezziages,valenti,ramirezlithium}). As is evident in Fig. \ref{fig:excess}, this fit underpredicts the NIR and mid-infrared photometry. One explanation for this would be that the star has an infrared excess due to hot, close-in dust. The best-fitting blackbody model for the excess with emissivity $\propto$ $\nu$ yields a dust temperature of 1465 $K$ and an infrared fractional luminosity ($L_{ir}/L_{*}$) of 0.01, both of which are quite high for such an old Sun-like star. However, \cite{ertelexcesses} found that \about 10$\%$ of Sun-like stars have bright exozodiacal dust, and the frequency of excesses may actually \textit{increase} with stellar age (for Sun-like stars); HD 11112A may fall into this category. Another possibility is that the dust resides around the imaged companion, which we reveal later to be a white dwarf.

\subsection{Companion Astrometry and Photometry}
\begin{figure}[h!]
\includegraphics[width=0.5\textwidth]{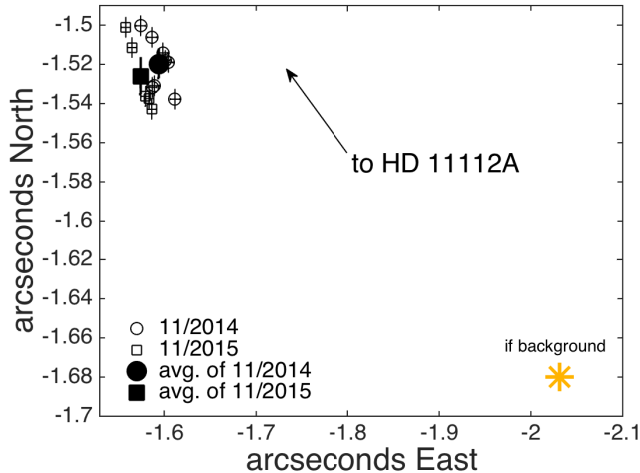}
\caption{Astrometry of the faint point-source (circles and squares) around HD 11112 obtained from MagAO's VisAO and Clio-2 cameras over the course of one year. The source is inconsistent with being a background object (yellow star) at more than 60$\sigma$ confidence. }
\label{fig:astrometry}
\end{figure}

Astrometry of the faint point-source was computed by calculating the photocenter inside circular apertures with radius = 0\fasec 1 (for VisAO images) or 0\fasec 08 (for Clio-2 images). Uncertainties were assumed to be 5 mas in the north and east directions, based on previous results with MagAO from \cite{me4796}. Fig. \ref{fig:astrometry} shows that, based on the primary star's high proper motion (0.415\asec east, 0.152\asec north, \citealt{updatedhip}) over the course of the year between the two imaging epochs, the faint companion is inconsistent with being a zero proper motion background object at more than 60$\sigma$ confidence. Therefore we consider the faint point-source to be a gravitationally-bound object and henceforth refer to it as HD 11112B. At a projected separation of \about 2\fasec 2 (99.7 AU), the companion does not show any statistically significant orbital motion over the course of one year. The final astrometry is reported in Table \ref{tab:phot}.

Calculating accurate photometry for HD 11112B was not a straightforward task considering that some images used NDs for calibration, some used unsaturated short exposures, and others were reduced using ADI+PCA, which introduces self-subtraction, over-subtraction, and other biases. In the following, we describe in detail the methods used to obtain photometry for each image in each epoch. 

At $r'$, the image quality is very sensitive to the observing conditions since the Strehl ratio is low (\about 10-30$\%$). Because of the unfavorable observing conditions during epoch 1, the first $r'$ image was not of high quality (characterized by a ``blobby," non-spherical PSF). Further complicating matters was the use of the ND, which at the time of the observations only had a few calibration measurements. To mitigate these concerns, when computing the photometry we used a very large circular aperture, with radius = 7 $\times$ the full-width half-maximum (FWHM) at $r'$ (radius = 0\fasec 14) to ensure that most of the flux from HD 11112B was included in the aperture, while still being as small as possible to maximize SNR. The flux inside the aperture was averaged, as was the flux inside the same aperture centered on the star in the unsaturated (ND) image. The stellar flux was then scaled up by the ND's diminution factor of 7176 $\pm$ 332. The $\Delta$ magnitude at $r'$ was computed by dividing the companion's flux by the scaled stellar flux. The uncertainty on the companion flux was computed by placing the same circular aperture at twelve equally-spaced azimuthal angles around the star at the same radius as the companion, averaging the fluxes inside these apertures, then computing the standard deviation of all the fluxes. The final uncertainty was the sum in quadrature of this uncertainty with the ND calibration uncertainty. All of the above steps were repeated for the epoch 2 $r'$ images. 

For $i'$, the procedure was identical to above, except an aperture with radius = 3 $\times$ the FWHM (radius = 0\fasec 07) was used and the ND scaling was 1317.99 $\pm$ 52.85. For $z'$, a 3$\times$FWHM (radius = 0\fasec 09) aperture was used and the ND scaling was not required. For $Ys$, a 3$\times$FWHM (radius = 0\fasec 10) aperture was used and the ND scaling was not required; in addition, to mitigate the effects of the bright ghost near the companion, we subtracted the average flux inside an annulus with inner radius = 3$\times$FWHM and outer radius = 6$\times$FWHM.

For the Clio-2 images, the procedure required that we account for the biases introduced by the ADI+PCA reductions. At each wavelength, a circular aperture with radius = 2$\times$FWHM (0.5$\times$FWHM for \lprime ~due to the lower SNR detection) was used to calculate the average flux of the companion. This flux was then scaled up by a correction factor, which was determined by inserting and recovering scaled down replicas of the unsaturated stellar PSF at twelve equally-spaced azimuthal angles around the star, calculating the average fluxes inside the same apertures centered on the recovered point-sources, then comparing these to the expected average flux inside the same aperture centered on the pre-inserted scaled-down point-source. Uncertainties were calculated as the sum in quadrature of the standard deviation of the average fluxes inside the same apertures placed at the twelve position angles (of the final image without any artificial sources inserted), and the standard deviation of the correction factors. 

\begin{figure*}[t]
\centering
\includegraphics[width=0.75\textwidth]{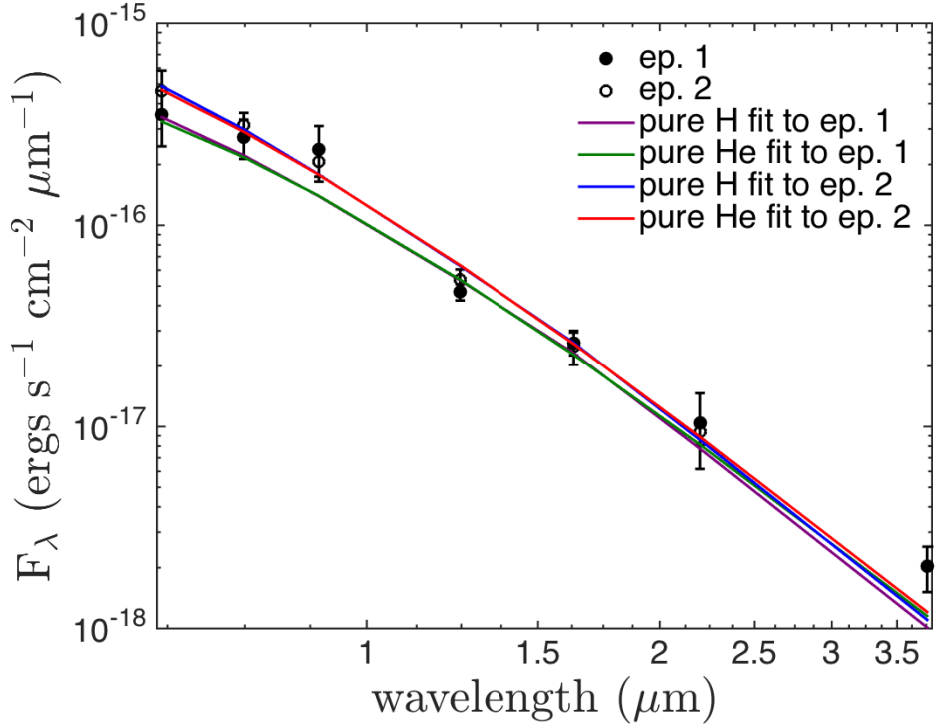}
\caption{The SED of HD 11112B from our MagAO images. The photometry over the two epochs is consistent within the errors. The colored lines are model fits to the data assuming pure H or pure He atmospheres of cool ($T_{eff} < 10,000 K$) white dwarfs with masses \about 0.9-1.1 \msun. }
\label{fig:SED}
\end{figure*}

To convert the $\Delta$ magnitudes into absolute magnitudes, we converted the catalog 2MASS photometry for HD 11112A into the MKO system using the color transformation relations in \cite{colortransforms} and used the derived catalog SDSS photometry for HD 11112A from \cite{ofek}, since the VisAO filters are very similar. We then added the $\Delta$ magnitudes to the primary's absolute magnitudes. We converted the absolute magnitudes into $F_{\lambda}$ (e.g. see \citealt{jackie13}) using the \textit{Hipparcos} parallax of 22.07$\pm$0.57 mas \citep{updatedhip}. All photometry is reported in Table \ref{tab:phot}, and the SED is shown in Fig. \ref{fig:SED}.

\begin{deluxetable}{c c c}
\tablecaption{HD 11112B Photometry and Astrometry \label{tab:phot}}
\tablehead{
\colhead{} & \colhead{Epoch 1} & \colhead{Epoch 2}
}
\startdata
%\hline \\
%\hline \\
Julian Date & 2456970.50000 & 2457356.50000 \\
$\Delta r'$ (0.63 \microns) & 10.23$^{+0.19}_{-0.23}$ & 9.94$^{+0.16}_{-0.19}$  \\
$\Delta i'$ (0.77 \microns) & 10.24$^{+0.14}_{-0.16}$ & 10.09$^{+0.09}_{-0.10}$  \\
$\Delta z'$ (0.91 \microns) & 10.04$^{+0.19}_{-0.23}$ & 10.19$^{+0.09}_{-0.10}$  \\
$\Delta Ys$ (0.99 \microns)\tablenotemark{a} & 9.68$^{+0.24}_{-0.32}$ &  -- \\
$\Delta J_{MKO}$ (1.1 \microns) & 11.02$^{+0.03}_{-0.03}$ & 10.87$^{+0.06}_{-0.06}$  \\
$\Delta H_{MKO}$ (1.65 \microns) & 10.85$^{+0.07}_{-0.07}$ & 10.89$^{+0.12}_{-0.13}$ \\
$\Delta Ks_{Barr}$ (2.15 \microns) & 10.81$^{+0.24}_{-0.31}$ & 10.91$^{+0.08}_{-0.09}$  \\
$\Delta$\lprime$_{MKO}$ (3.76 \microns) & 10.43$^{+0.16}_{-0.19}$ & --  \\
$M_{r'}$ & 13.99 $\pm$ 0.29 & 13.70 $\pm$ 0.25  \\ 
$M_{i'}$ & 13.83 $\pm$ 0.22 & 13.69 $\pm$ 0.15 \\
$M_{z'}$ & 13.59 $\pm$ 0.29 & 13.74 $\pm$ 0.16  \\
$M_J$ & 13.73 $\pm$ 0.10 & 13.57 $\pm$ 0.12 \\
$M_H$ & 13.34 $\pm$ 0.13 & 13.38 $\pm$ 0.19 \\
$M_{K}$ & 13.17 $\pm$ 0.37 & 13.27 $\pm$ 0.15 \\
$M_{L^{\prime}}$ & 12.75 $\pm$ 0.24 & -- \\
\hline 
$\Delta R.A.$ (\asec) & -1.59$\pm 0.01$ & -1.58$\pm 0.01$ \\
$\Delta Decl.$ (\asec) & -1.52$\pm 0.01$ & -1.53$\pm 0.01$  \\
$\rho$ (\asec) & 2.20$\pm 0.01$ & 2.20$\pm 0.01$ \\
$P.A.$ (\degrees) & 226.4$\pm 0.2$ & 225.9$\pm 0.2$ \\
\enddata
\tablenotetext{a}{$Ys$ data are not used in the SED modeling.}
%\tablecomments{blah}
\end{deluxetable}

\subsection{SED Fitting}
\label{sec:sed}
While the NIR colors of HD 11112B point to it possibly being a cool brown dwarf \citep{leggettbrowndwarfs}, the fact that it is bright at optical wavelengths suggests instead that it is a white dwarf (e.g., \citealt{crepptrends3}). Therefore we set out to fit the companion's photometry to cool white dwarf models. While the quality of the epoch 1 data was poor, we included these data in the analysis for completeness (and the final fitting results are not markedly different compared to epoch 2). For cool white dwarfs, the atmospheric parameters ($T_{eff}$ and $\log g$) and chemical compositions can be measured accurately using the photometric technique developed by \citet{BRL97}. To fit the SED of HD 11112B, we first converted the optical and infrared photometric measurements into observed fluxes using the procedure outlined in \citet{holberg06}, including the zero points for the various photometric systems. The transmission functions in the optical for the $ugriz$ photometry is described in \citet{holberg06} and references therein, while those for the Mauna Kea Observatories (MKO) photometric systems are taken from \citet{MKO}.

We then related the average observed fluxes $f_{\lambda}$ and the average model fluxes $H_{\lambda}$ --- which depend on $T_{eff}$, $\log g$, and chemical composition --- by the equation
\begin{equation}
f_{\lambda} = 4\pi(R/d)^2 H_{\lambda}, 
\end{equation}
where $R/d$ defines the ratio of the radius of the white dwarf to its distance from Earth. Next we minimized the $\chi^2$ value defined as the difference between observed and model fluxes over all bandpasses with weights determined by the photometric uncertainties. We used the nonlinear least-squares method of Levenberg-Marquardt \citep{press86}, which is based on a steepest descent method. Only $T_{eff}$ and the solid angle $\pi (R/d)^2$ were considered free parameters, while the uncertainties of both parameters were obtained directly from the covariance matrix of the fit. We started with $\log g=8$ and determined the corresponding effective temperature and solid angle, which combined with the distance, $d$ (obtained from the known parallax) gives us the radius of the white dwarf, $R$. We then converted the radius into mass using evolutionary models similar to those described in \citet{fon01} but with C/O cores: $q({\rm He})\equiv \log M_{\rm He}/M_{\star}=10^{-2}$ and $q({\rm H})=10^{-4}$, which correspond to hydrogen-atmosphere white dwarfs, and $q({\rm He})=10^{-2}$ and $q({\rm H})=10^{-10}$, which correspond to helium-atmosphere white dwarfs. Since the atmospheric composition of HD 11112B is unknown, we assumed pure hydrogen and pure helium model atmospheres and used the synthetic spectra described in \citet{bergeron95}, \citet{TB09}, \citet{bergeron11}, and references therein. In practice, the $\log g$ value obtained from the inferred mass and radius ($g=GM/R^2$) is different from our initial guess of $\log g=8$. Therefore the fitting procedure was repeated until an internal consistency in $\log g$ was reached.
\begin{deluxetable*}{ccccccc}
\tablecaption{HD 11112B SED Fitting Results \label{tab:models}}
\tablehead{
\colhead{} & & \multicolumn{2}{c}{Epoch 1} & & \multicolumn{2}{c}{Epoch 2} \\
\cline{3-4}\cline{6-7} \\
\colhead{} & & \colhead{Pure H} & \colhead{Pure He} & & \colhead{Pure H}  & \colhead{Pure He} 
}
\startdata
\reducedchi & &  2.47 & 2.37 & & 1.62 & 1.47 \\
Mass (\msun) & & 1.06$^{+0.02}_{-0.02}$ & 0.90$^{+0.021}_{-0.02}$ &  &1.08$^{+0.02}_{-0.02}$ & 0.96$^{+0.02}_{-0.02}$ \\
$T_{eff}$ ($K$) & & 8400$^{+2000}_{-2000}$ & 7300$^{+1900}_{-1900}$ & & 9800$^{+1700}_{-1700}$ & 8700$^{+1800}_{-1800}$ \\
$\log{g}$ (cm s$^{-2}$) & & 8.73$^{+0.03}_{-0.03}$ & 8.50$^{+0.03}_{-0.03}$ & & 8.77$^{+0.03}_{-0.03}$ & 8.59$^{+0.03}_{-0.03}$ \\
$\log{L/L_{\odot}}$ & & -3.61$^{+0.02}_{-0.02}$ & -3.69$^{+0.02}_{-0.02}$ & & -3.39$^{+0.03}_{-0.02}$ & -3.46$^{+0.02}_{-0.02}$ \\
Age$_{cool, 50\% C/O}$ (Gyr)\tablenotemark{a} & & 3.17$^{+1.90}_{-1.27}$ & 3.53$^{+0.92}_{-1.41}$ & & 2.43$^{+1.03}_{-0.70}$ & 2.88$^{+1.02}_{-1.22}$ \\
Age$_{cool, 100\% C}$ (Gyr)\tablenotemark{b} & & 3.58$^{+2.38}_{-1.63}$ & 3.92$^{+1.45}_{-1.61}$ & & 2.65$^{+1.43}_{-1.04}$ & 3.15$^{+1.28}_{-1.54}$ \\
\enddata
\tablenotetext{a}{The cooling age of the white dwarf, assuming a 50$\%$ C/O core and taking into account the uncertainties in mass and temperature.}
\tablenotetext{b}{The same, but for a 100$\%$ C core.}
%\tablecomments{blah}
\end{deluxetable*}

The best-fitting models for the epoch 1 and epoch 2 photometry are shown in Fig. \ref{fig:SED} as well as in Table \ref{tab:models}. The reduced $\chi^2$ (\reducedchi) ranges from \about 1.45-2.5, indicating overall good fits. Interestingly, the \lprime ~flux is 1.6-2$\sigma$ larger than the models predict. Pure He atmospheres provide marginally better fits than pure H. The estimated white dwarf masses and effective temperatures range from \about 0.9-1.1 \msun ~and 7300-9750 $K$, respectively. These correspond to progenitor masses ranging from 4.3-6.5 \msun ~\citep{wdprogenitormasses,williamswdprogenitors}, which have main sequence lifetimes $\lesssim$ 160 Myr \citep{massivestarages}. This is insignificant compared to the expected cooling age for the white dwarf, which we modeled explicitly. 

Determining the cooling age of a white dwarf by fitting its observed mass, atmospheric composition and temperature using evolutionary sequences requires making an assumption about its core composition; the effect of this assumption is most pronounced in the case of older white dwarfs \citep{fon01}. As the thermonuclear burning rate of He is uncertain, the exact core compositions of white dwarfs are generally unknown. Attempts to obtain such measurements have shown that white dwarf cores must at least be partly composed of oxygen and are perhaps even dominated by it (e.g., see \citealt{salaris,althaus}). More recently, \cite{fields} performed a Monte Carlo simulation to estimate the core composition of a 0.64 \msun ~white dwarf using the STARLIB reaction rate library and the MESA evolutionary code. It is the first time that such an analysis accounted for uncertainties in the $^{12}C(\alpha, \gamma)^{16}O$, the triple-$\alpha$, and the $^{14}N(\iota{p},\gamma)^{15}O$ nuclear reaction rates. They found that it is practically impossible to precisely infer the core composition of white dwarfs given the current uncertainties on the best available measurements for the aforementioned nuclear reaction rates. However, they were able to show that the core compositions of their simulated white dwarfs were of at least 25$\%$ oxygen at 95$\%$ confidence. \cite{whitedwarfcomposition} have produced the only reliable direct measurement of the core composition of a white dwarf to date, using asteroseismology to deduce that the 0.65 \msun ~white dwarf Ross 548 has a fractional oxygen core composition of $X$(O)\,=\,0.70\,$\pm$\,0.06. 

Based on these previous results, for HD 11112B we assumed a core composition of 50$\%$ C and 50$\%$ O and used the evolutionary models described in \cite{BLR01}. The resulting cooling ages ranged from \about 2.4-3.5 Gyr. If we assume an unrealistic 100$\%$ C core, then the cooling ages are slightly larger, ranging from \about 2.6-4 Gyr. 

\subsection{Constraints from RV Analysis}
The RVs for HD 11112 show no statistically significant curvature, therefore it is difficult to estimate the period of the companion HD 11112B. However, its mass can be estimated following the procedure outlined in \cite{rvimagingconstraints} and \cite{milo1}. Briefly, we used a Bayesian Markov chain Monte Carlo (MCMC) approach to produce posterior distributions of the allowed parameter values \citep{fordmcmc}. The likelihood function $L$, which contains the Keplerian model and nuisance parameters, is simpler in this case because data were obtained with only one instrument and there are no obvious planetary signals in the data (just a long-period linear trend). The likelihood function is given by
\begin{eqnarray}
%L &=& \prod_{I}\prod_i^{N_{obs}} l_{i,I}\\
L &=& \frac{1}{\sqrt{2\pi}}\frac{1}{\sqrt{\epsilon_{i}^2+ s^2}}
\exp\left[
    -\, \frac{1}{2} \frac{\left(v_{i}-v(t)\right)^2}{
         \epsilon_{i}^2+ s^2
	 }
\right]\\
v_{i} &=& \gamma + \sum_p m(\hat{\kappa}_p; t) + \dot{v}_r(t-t_0),\label{eq:rvmodel}
\end{eqnarray}
where $i$ indexes the individual observations, $\epsilon_{i}$ is the nominal uncertainty of each RV measurement, $\gamma$ and $s$ are the zero-point and extra noise parameters (also called jitter), and the Doppler signal from a companion on the star is encoded in the model $ m(\hat{\kappa}_p; t)$, which is a function of time $t$ and the Keplerian parameters $\hat{\kappa_p}$. The Keplerian parameters of the p$^{th}$ companion in the system are: the orbital period $P_p$ (in days), the semi-amplitude $K_p$ (in m/s), the mean anomaly $\mu_{0,p}$ at the reference epoch $t_0$ (in degrees), the eccentricity $e_p$, and the argument of periastron $\omega_p$. The second term in Eq. \ref{eq:rvmodel} accounts for the possible presence of a long-period candidate whose orbit is only detected as a trend (acceleration, $\dot{v}_r$). A third term, (jerk, $\ddot{v}_r$) was initially included, but fits to the data showed that there was no statistically significant curvature, so this term was later dropped. Additional details on our MCMC fitting can be found in Section 3.3 of \cite{milo1}.

\begin{figure}[t]
\centering
\includegraphics[width=0.5\textwidth]{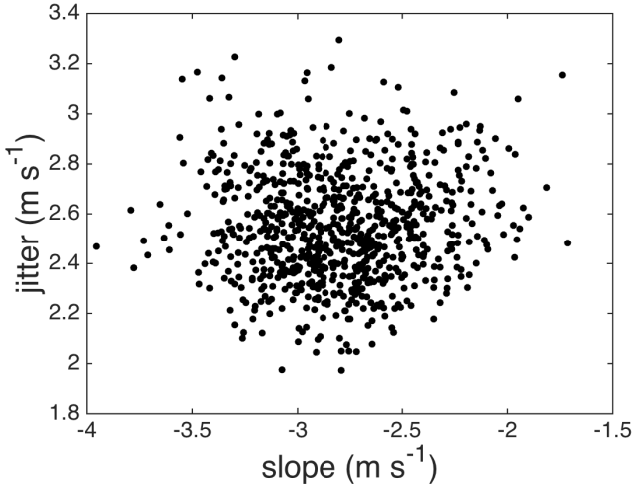}
\caption{Slope vs. jitter values from our MCMC analysis of the RV data. The slope terms are clustered around -2.8 \ms, while the jitter values are \about 2.5 \ms, fully consistent with expectations for an early G star \citep{jitter}. }
\label{fig:jitter}
\end{figure}

For HD 11112, initial periodogram results for a single planet showed that there was a peak at 1.46 days with a false alarm probability of 5$\%$, making the signal dubious. Therefore the planetary term (the second term in Eq. \ref{eq:rvmodel}) was dropped, leaving only the long-period term (the slope). Fits to the data were then computed. The significance of the slope (median = -2.82 \ms) being nonzero was 8$\sigma$ (see Fig. \ref{fig:jitter}). We used the fitted slope terms to compute the posterior mass distribution of HD 11112B using

\begin{eqnarray}
\frac{M_B}{M_\odot} &=&  5.341 \times 10^{-6} \,\, \dot{v}_r
\,\,\left(\frac{\rho}{\Pi}\right)^{2} \frac{1}{\Psi} \,\,,       \label{eq:ftorres}       \\
\end{eqnarray}
where $M_B$ is the mass of HD 11112B in solar masses, $\dot{v}_r$ is the slope term generated by the MCMC procedure, $\rho$ is the current projected separation of HD 11112B, $\Pi$ is the parallax, and $\Psi$ contains the angle and time terms:
\begin{eqnarray}
\Psi &=& [(1-e)(1+\cos E)]^{-1} (1-e \cos E) \sin i \times \\
 & & (1-\sin^2(\nu+\omega)\sin^2 i) (1+\cos\nu)\sin(\nu+\omega) 
\,
\label{eq:psi}
\end{eqnarray} 
where $E$ is the eccentric anomaly, $i$ is the inclination, $\nu$ is the true anomaly, and $\omega$ is the argument of periastron.
\begin{figure*}[t!]
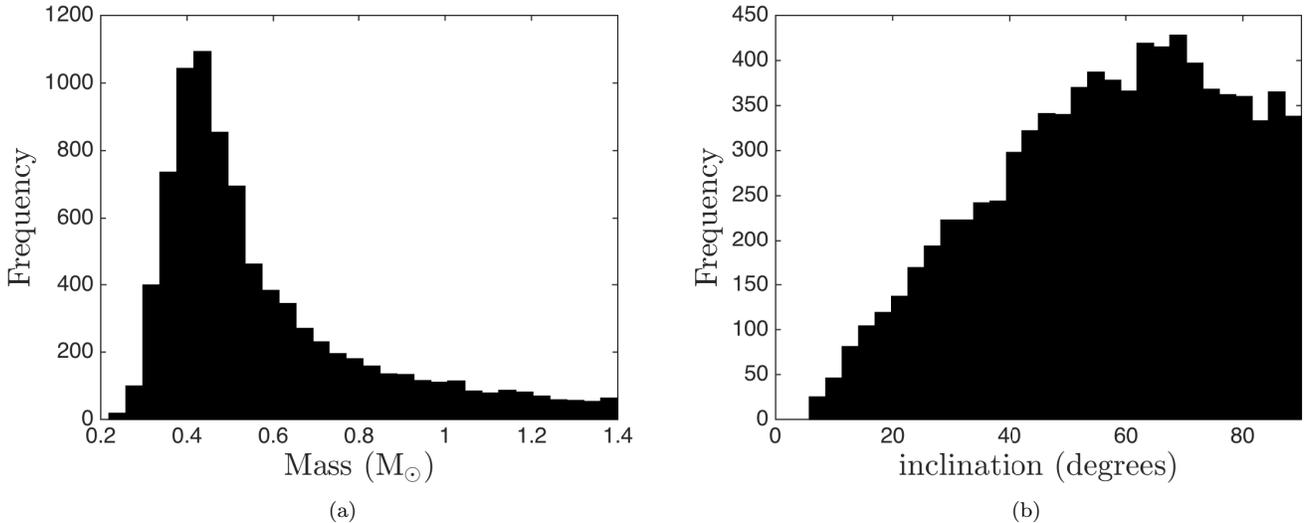

\gridline{\fig{mass.eps}{0.5\textwidth}{(a)}
		\fig{inc.eps}{0.5\textwidth}{(b)}
         } 
%\gridline{\fig{massvsinc.eps}{0.5\textwidth}{(c)}
	%	\fig{massvse.eps}{0.5\textwidth}{(d)}
     %    } 
\caption{Posterior distributions for the mass (a) and inclination (b) of HD 11112B. After assuming a mass cut-off of 1.4 \msun (the Chandrasekhar limit), at 99$\%$ confidence its minimum mass is 0.28 \msun ~and its minimum inclination is 9.8\degrees.}
\label{fig:dist}
\end{figure*}

\begin{figure*}[t!]
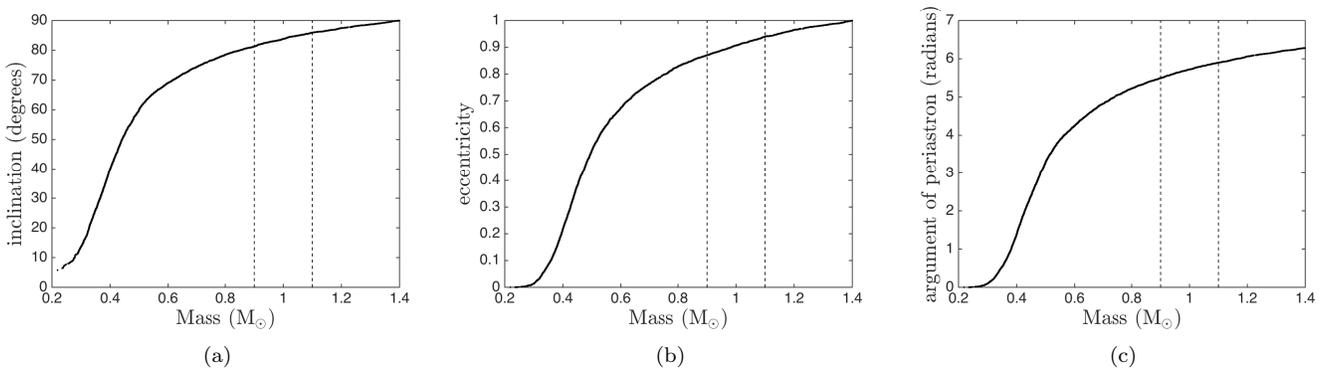

\gridline{\fig{massvsinc.eps}{0.33\textwidth}{(a)}
		\fig{massvse.eps}{0.33\textwidth}{(b)}
		\fig{massvsw.eps}{0.33\textwidth}{(c)}
         } 
\caption{Mass vs. inclination (a), mass vs. eccentricity (b), and mass vs. argument of periastron (c) for HD 11112B. The nominal SED-fit mass range (0.9-1.1 \msun) is denoted by the dashed lines. These high masses correspond to near edge-on, high-eccentricity orbits for the white dwarf companion.}
\label{fig:massplots}
\end{figure*}

We generated the posterior distribution for the mass of HD 11112B using Eq. \ref{eq:ftorres} while randomly drawing from Gaussian distributions centered on the measured values and including the corresponding uncertainties for $\Pi$ and $\rho$, and also randomly drawing from uniform distributions for the other variables ($\cos{i}$ and $e$ between 0-1, $\omega$ between 0-2$\pi$). The resulting distribution, shown in Fig. \ref{fig:dist}, is sharply peaked at \about 0.42 \msun. Given the high likelihood that HD 11112B is a white dwarf (see Section \ref{sec:sed}), we imposed a mass cutoff of 1.4 \msun (the Chandrasekhar limit; \citealt{wdmasslimit}). Based on this constraint, at 99$\%$ confidence the minimum mass of HD 11112B is 0.28 \msun ~and the median mass is 0.49 \msun. We can also examine the resulting posterior distributions for the inclination (top right panel of Fig. \ref{fig:dist}), finding at 99$\%$ confidence a minimum inclination of 9.8\degrees. The other orbital parameters were fully unconstrained; however, some inferences can be gleaned by examining the relationship between the mass, inclination, eccentricity, and argument of periastron (Fig. \ref{fig:massplots} bottom panels). For the best-fitting SED mass range (0.9-1.1 \msun), the white dwarf companion should have a near edge-on, high-eccentricity orbit. 

We can also make some inferences on its likely semimajor axis ($a$) using the information at hand. Since the object is most likely near apoastron, $r$ \about $r_{apo} = a (1 + e)$, where $r_{apo}$ is the apoastron distance. We also know that the object's current projected separation is the closest it could be to the primary, or $r > r_{proj}$. Therefore, assuming $e$ \about 0.9 from Fig. \ref{fig:massplots}, $a \gtrsim $ 50 AU (period longer than \about 300 years). This means that we should not expect much orbital motion over the next few years.

%\begin{figure}[t!]
%\includegraphics[width=0.5\textwidth]{mass.eps}
%\caption{blah}
%\label{fig:mass}
%\end{figure}

\section{Discussion: A Puzzling White Dwarf}
In this work, we have shown that the HD 11112 system is a binary consisting of a Sun-like evolving G dwarf and a secondary white dwarf. SED modeling suggests that the white dwarf is cool ($T_{eff} < 10,000 K$) and has a mass of \about 0.9-1.1 \msun. These physical properties correspond to cooling ages ranging from \about 2.4-4 Gyr (Table \ref{tab:models}). 

The SED mass falls in the tail of the posterior mass distribution from our RV analysis, corresponding to a 25$\%$ chance the mass is $>$ 0.9 \msun. However, white dwarf models have been shown to be robust and have been calibrated on objects  like HD 11112B with accurate parallaxes (e.g., see \citealt{wdcalibrations}). Therefore it seems very plausible that we have found a rather unusual high-mass white dwarf, which is statistically rare in and of itself (white dwarf mass distribution in the solar neighborhood being peaked at \about 0.6-0.7 \msun; \citealt{wdcalibrations,wdmasses,nearbywhitedwarfs}). In addition, we have to reconcile the apparent age discrepancy with the primary star (age = 7.2$^{+0.78}_{-1.2}$ Gyr).

Assuming a 50$\%$ C/O core composition, the white dwarf cooling age is at best $2.4\sigma$ smaller than the primary's age. The only way to reconcile this discrepancy is to assume an (unlikely, see Section \ref{sec:sed}) 100$\%$ C core. In this case, the cooling age is 3.58$^{+2.38}_{-1.63}$ Gyr and marginally consistent at the 1.4$\sigma$ level. However, this corresponds to model fits to the epoch 1 data, which were of much poorer quality than the epoch 2 data. If we restrict ourselves to the epoch 2 data alone, then for a 100$\%$ C core, the age discrepancy is at best at the 2.3$\sigma$ level. 

One way to reconcile the age discrepancy is if there was a \textit{delay} in HD 11112B's evolution to the white dwarf phase. This could be achieved if HD 11112B was originally a close binary (and the HD 11112 system was therefore a hierarchical triple system). The two stars could have spent several Gyr on the main sequence and then either (1) merged into a single, high-mass blue straggler that then evolved into the observed white dwarf or (2) evolved separately into two low-mass white dwarfs that then merged into the observed high-mass white dwarf. An example of such a system was recently discovered by \cite{wdmerger}, where the ``delayed" white dwarf has a final mass of \about 0.85 \msun. 

We can infer some properties of the binary progenitors based on the age constraints from the primary and the observed white dwarf. The total age of the system is \about 7 Gyr (from the primary), and the cooling age of the white dwarf (assumed to have mass of \about 1 \msun) is at most \about 4 Gyr. The white dwarf progenitor would have a mass of \about 5 \msun ~\citep{williamswdprogenitors} and live on the main sequence for \about 125 Myr \citep{massivestarages}. Therefore the process that produced the white dwarf progenitor has \about 2.9 Gyr of evolution to account for (if the progenitor is a single star; otherwise 3 Gyr to account for). During a merger of two main sequence stars, only a few percent of the total input mass is lost \citep{mergermass}. Thus we can take the white dwarf single-star progenitor mass as an upper limit on the total pre-merger mass. If the two stars in the binary are identical, they would have masses \about 2.5 \msun ~and each live on the main sequence for \about 765 Myr, far short of the required 2.9 Gyr. In fact, in order for the two identical main sequence stars to merge after 2.9 Gyr, the pair would have to each be \about 1.6 \msun ~for a total of 3.2 \msun, which is far short of the expected 5 \msun ~white dwarf progenitor mass. 

We are thus left with three possible scenarios. (1) one star in the binary has mass $\lesssim$ 1.6 \msun, the other star is more massive and evolves into a white dwarf first, and then the white dwarf merges and is absorbed into the other star after \about 3 Gyr. Unfortunately, the total merged mass (even for a white dwarf with mass = 1.4 \msun) would still fall short of the required 5 \msun ~progenitor, so this scenario seems unlikely. (2) The same formation happens as in (1), except that the white dwarf accretes material from the lower-mass main sequence star after it evolves off the main sequence. The binary would become a cataclysmic variable whose final fate could be completely self-destructive, so this scenario seems unfavorable. (3) Both stars in the binary evolve into white dwarfs and then merge into a more massive white dwarf. While white dwarf mergers often result in supernova explosions \citep{wdmergerfates}, two low-mass (total mass $<$ 1.4 \msun) white dwarfs can merge into a more massive white dwarf as long as dynamic carbon burning does not occur during the merger phase \citep{massivewdmerger}. In fact, this is the favored scenario to explain most of the massive white dwarfs in the solar neighborhood \citep{nearbywhitedwarfs}. For HD 11112, the timing works out as long as the two white dwarfs each had masses $\lesssim$ 0.55 \msun, which correspond to progenitor main sequence lifetimes of \about 3 Gyr. In order to evolve into two white dwarfs and merge in this timeframe, the binary would have to shrink to an orbital period of \about 5 hours. This would naturally happen if the pair had undergone common envelope evolution due to dynamical friction with Roche lobe material from both stars. After reaching such a small orbit, it would continue to decay and merge in \about 3 Gyr \citep{orbitaldecay}. This seems like the most plausible explanation for the peculiarities of HD 11112B.

This would appear to resolve the puzzling nature of HD 11112B. The only other similar benchmark white dwarf (HD 114174B, likewise detected by both RV and direct imaging, \citealt{crepptrends3}), is also discrepant with its primary star's age \citep{lbtwd}. In this case, the white dwarf cooling age is actually \textit{larger} than the primary's age and so may be more difficult to explain. Benchmark objects like HD 114174B and HD 11112B are perhaps the best candidates for testing white dwarf models because they have been resolved, they have measured ages via their primaries, and their orbital motions and their RVs can be used to constrain their masses with continued monitoring over time. 

%Finally, HD 11112B is interesting because it has its own marginal infrared excess at 4 \microns ~(Fig. \ref{fig:SED}. If verified, this would be especially intriguing given that the puzzling nature of the primary's infrared photometry, one explanation for which is very hot, close-in dust (Section \ref{sec:stellar}). The excess around the companion needs to be confirmed with additional infrared photometric measurements. Unfortunately, the white dwarf is too close to the primary star to be resolvable by $WISE$ \citep{wise} or $Spitzer$ \citep{spitzer}, both of which have optimal resolutions and pixel scales that are too large (\about a few arcseconds). HD 11112B, however, would be a good target for $JWST$'s $MIRI$ instrument \citep{MIRI}, which will be able to obtain high-resolution images from 5-27 \microns. $MIRI$ could thus map out the mid-infrared SED of the white dwarf and confirm/refute the infrared excess. 

Given the intriguing nature of HD 11112B, the HD 11112 system warrants further study. At $>$ 2\asec ~separation, \textit{GAIA} should provide high-quality astrometric data to help refine the orbit \citep{gaiaplanets}, though the object should be moving very slowly due to its likely large orbit and high eccentricity (Fig. \ref{fig:massplots}). The companion should also be easily detected by extreme AO systems like GPI \citep{gpi} and SPHERE \citep{sphere}, which would help not only with astrometric and photometric monitoring, but also potentially with finer characterization of the object via spectroscopy or polarization. 

% age discrepancy
% infrared excess

\acknowledgments
This work was supported in part by the NSERC Canada and by the Fund FRQ-NT (Qu{\'e}bec). T.J.R. acknowledges support for Program number HST-HF2–51366.001-A, provided by NASA through a Hubble Fellowship grant from the Space Telescope Science Institute, which is operated by the Association of Universities for Research in Astronomy, Incorporated, under NASA contract NAS5-26555. This publication makes use of data products from the Two Micron All Sky Survey, which is a joint project of the University of Massachusetts and the Infrared Processing and Analysis Center/California Institute of Technology, funded by the National Aeronautics and Space Administration and the National Science Foundation. This publication also makes use of data products from the Wide-field Infrared Survey Explorer, which is a joint project of the University of California, Los Angeles, and the Jet Propulsion Laboratory/California Institute of Technology, funded by the National Aeronautics and Space Administration.

\facilities{Magellan-Clay\footnote{This paper includes data obtained at the 6.5 m Magellan Telescopes located at Las Campanas Observatory, Chile.}, Anglo-Australian Telescope}

\bibliographystyle{aasjournal}
\bibliography{wd}

\end{document}